\documentclass[aps,12pt,final,notitlepage,oneside,onecolumn,nobibnotes,nofootinbib,%
superscriptaddress,noshowpacs,centertags]{revtex4}
\usepackage{amsmath,amssymb,amsfonts,amsbsy}
\usepackage{xcolor}
\usepackage{graphicx}
\usepackage{multirow}
\usepackage[normalem]{ulem} 

\begin{document}

\title{Rydberg states of muonic helium in quantum electrodynamics}
\author{\firstname{A.~V.}~\surname{Eskin}}
\affiliation{Samara University, Samara, Russia}
\author{\firstname{A.~P.}~\surname{Martynenko}}
\affiliation{Samara University, Samara, Russia}
\author{\firstname{F.~A.}~\surname{Martynenko}}
\affiliation{Samara University, Samara, Russia}
\author{\firstname{D.~K.}~\surname{Pometko}}
\affiliation{Samara University, Samara, Russia}

\begin{abstract}
The variational method is used to study the energy levels of muonic helium $(\mu^--e^--He)$ with an 
electron in the ground state and a muon in an excited state with principal and orbital quantum numbers 
$n\sim\ l+1\sim 14$ (the Rydberg muon states). The variational wave functions are chosen 
in the Gaussian form. The matrix elements 
of the Hamiltonian in the nonrelativistic approximation, as well as corrections for the vacuum polarization 
and relativism, are calculated analytically. A series of energies of the 
high-angular-momentum muons states is obtained, which can be studied experimentally.
\end{abstract}



\maketitle

\section{Introduction}
\label{vv}

The Rydberg states in atoms play an important role in refining the values of fundamental constants. For example, 
measurements of the Rydberg constant \cite{H1,H2} have been improved using the Rydberg state spectroscopy 
in the hydrogen atom. Working with the Rydberg states can reduce the influence of strong interaction corrections 
on the energy level structure.
The electron Rydberg atoms are characterized by very low ionization energies and have characteristic sizes that 
increase as $n^2$ \cite{RA} compared to the ground state size. Since the radiative lifetime of the excited state 
is proportional to $n^3$, the Rydberg atoms are characterized by extremely long lifetimes: from a thousandth 
of a second to one second. 
Using the method of stepwise photoexcitation with tunable lasers, the Rydberg atoms with principal quantum numbers 
$n\sim 1000$ have been obtained. The low density of matter in the interstellar medium and, consequently, the low probability of collisional ionization, determine the long lifetimes of the Rydberg atoms in space. Electromagnetic 
radiation from the Rydberg atoms makes it possible to determine the composition and characteristics of a matter 
in the interstellar medium. Thus, the Rydberg atoms of hydrogen with $n > 150$ and carbon with $n > 700$ 
were discovered using radiation lines in the radio range in outer space.

In the paper \cite{merkt}, the Rydberg levels of the hydrogen atom
in an electric field were studied. The Lamb shift-based measurement of the proton charge radius $r_p$ in muonic 
hydrogen atom yields a smaller value than the measurement of $r_p$ in ordinary hydrogen. This discrepancy 
is known as the "proton radius puzzle" \cite{aldo,uj}. $r_p$ is found indirectly from measurements 
of the Rydberg constant $R$, where the electron wave function partially overlaps with the proton volume, 
and a correlation between $r_p$ and $R$ should exist. Therefore, experiments with highly excited Rydberg states, 
which are sensitive to $R$ but have little effect on $r_p$, are important for resolving the $r_p$ problem. 
However, such measurements would require very high frequency resolution. In the work \cite{merkt}, this difficulty 
was overcome by studying hydrogen in an electric field. Transitions 
to the Rydberg states (with a known correction for the Stark effect) with quantum numbers $n=20$ and $n=24$ were measured, and the ionization energy was found. The Rydberg frequency $cR$ was obtained using a method insensitive to $r_p$. The result \cite{merkt} confirms the value of $r_p$ obtained in the experiment with muonic hydrogen and contributes to a more accurate resolution of the $r_p$ problem. It can be said quite definitely that, along with the study of low-lying energy levels \cite{egs} in the hydrogen atom, the study of highly excited states is of considerable interest for a more precise determination of a number of physical parameters of the theory with high accuracy.

The aim of this work is to study excited states of high-angular-momentum muons in muon-electron helium 
atoms, in which the muon and electron are approximately equidistant from the nucleus. From here on, 
we refer to such states as the Rydberg muon states, emphasizing that the principal quantum number 
for the muon orbit is quite large.
When a beam of negative muons enters a helium target, it is slowed to a kinetic energy of several 
tens of eV due to ionization and excitation 
of target molecules. An exotic muonic helium atom is formed when a muon stops within the target and is captured 
by the target atom into an outer atomic orbit, replacing one of the electrons in an orbit with $n\leq 14$. The capture 
is a complex process and depends on the atomic and molecular structure, the range of values of the principal 
quantum number n, and the angular momentum l. The lifetime of a muonic helium atom is determined 
by large orbital angular momentum of $l = (10\div 13)$, at which point the muon is formed. 
Its transition to the ground state with $l = 0$ is strongly suppressed. The lifetime of such an atom is a few nanoseconds. After capture, the muon cascades from its high-angular-momentum state 
through the atom's own series of bound states, located within the electron levels due to its greater mass. 
The cascade is accompanied by the emission of X-rays and ends when the particle is absorbed 
by the nucleus as a result of the weak interaction. The muon cascades downward through the Coulomb deexcitation 
and external Auger emission until the probability of radiative transitions becomes dominant for lower-lying states.
The cascade time in an exotic muonic helium atom is short compared to the particle lifetime, making X-ray processes observable using spectroscopic methods. In the case of muonic Rydberg states in muonic helium, the influence 
of nuclear structure effects on the emission spectrum is significantly reduced. For pionic helium atoms,
studies of the Rydberg states allow us to obtain a more accurate pion mass than is known from other sources 
\cite{hori1,hori2,bakalov}.
M. Hori's group plans to perform laser spectroscopy of the kaonic atom using the kaon beam at DAPHNE, which 
will improve the accuracy of charged kaon mass determination by 2-3 orders of magnitude compared to the current 
level.

It is worth noting that other approaches to determining the $\pi$-meson mass exist. A study conducted 
in \cite{crystal} demonstrates the potential of crystal spectroscopy of exotic atoms. In this work, the $5g-4f$ transitions in pion nitrogen and muon oxygen were measured simultaneously in a gaseous nitrogen-oxygen mixture. 
Knowing the muon mass, the muon line can be used to energy-calibrate the pion transition. The obtained mass of the negatively charged pion is 4.2 ppm higher than the current world average of $139.57077\pm 0.00017$ MeV \cite{pdg}.

The energy levels of three-particle systems can be studied with high accuracy using the variational method. 
There are some differences in the use of the variational method for finding the energy levels of three-particle 
systems. These differences relate to the choice of coordinates and the representation of the Hamiltonian to describe 
the system, as well as the choice of basis wave functions. Thus, in the work \cite{hori1}, an exponential basis 
was used, and the coordinates of the electron and muon were determined relative to the nucleus. In the works \cite{phys2023,apm2026}, when calculating the energy levels of hydrogen mesomolecules, muonic helium, 
and other molecules, we use the formulation of the variational method in the Jacobi coordinates with the Gaussian wave 
functions in the case of theree-particle and four-particle systems. 
Using this approach, we calculated the energy levels of the Rydberg states in pion and kaonic helium atoms \cite{apm2024,apm2024a}. In this paper, we extend the scope of research to muonic helium atoms $(\mu^--e^--He)$ 
and perform energy level calculations for states in which 
the muon is in an excited state with a large orbital quantum number $l=10, 11, 12, 13$.

\section{General formalism}
\label{gf}

Various approaches have been developed to study the energy levels of three-particle systems. 
One method, analytical perturbation theory, allows for analytical investigation of both the Lamb shift 
and the hyperfine structure of the spectrum \cite{mohr,huang,amusia,amusia1,apm2008,apm2022,apm2021}.
Other methods used for many-particle systems
are the variational method and the hyperspherical coordinate method, which allow for very high-precision 
determination of the energy levels and wave functions of bound states
\cite{rd,melezhik,puchalski1,puchalski2,frolov,frolov1,chen,iran,varga,korobov,khan}. 
Since the muonic helium system considers 
atomic states with large muon orbital angular momenta, where the electron and muon are equidistant 
from the nucleus, 
it is virtually impossible to use analytical perturbation theory. Therefore, we further study this system 
using the variational method. Gaussian wave functions are used as basis wave functions.

To find the energy levels of a three-particle system, we introduce the Jacobi coordinates $\boldsymbol\rho$,
$\boldsymbol\lambda$, connected with the radius vectors of the particles ${\bf r}_1$ (nucleus), ${\bf r}_2$ 
(muon), ${\bf r}_3$
(electron) as follows:
\begin{equation}
{\boldsymbol\rho}={\bf r}_2-{\bf r}_1,~~~{\boldsymbol\lambda}={\bf r}_3-\frac{m_1{\bf r}_1+m_2{\bf r}_2}{m_1+m_2},
\label{eq1}
\end{equation}
where $m_1$, $m_2$, $m_3$ are the masses of the $He$ nucleus, muon and electron.

To solve the variational problem, we choose the wave functions of the ground state in the form 
of a superposition of Gaussian exponentials:
\begin{equation}
\label{eq2}
\Psi({\boldsymbol\rho},{\boldsymbol\lambda},A)=\sum_{i=1}^K C_i \psi_i({\boldsymbol\rho},{\boldsymbol\lambda},A^i),
~~~\psi_i({\boldsymbol\rho},{\boldsymbol\lambda},A^i)=e^{-\frac{1}{2}\left(A_{11}^i{\boldsymbol\rho}^2+
2A_{12}^i{\boldsymbol\rho}{\boldsymbol\lambda}+A_{22}^i{\boldsymbol\lambda}^2\right)},
\end{equation}
where $C_i$ are the linear variational parameters, $A^i$ is the matrix of nonlinear variational parameters, 
K is the size of the basis.

In the nonrelativistic approximation, the Hamiltonian of a three-particle atom in Jacobi coordinates 
can be represented as follows:
\begin{equation}
\hat H_0=-\frac{1}{2\mu_1}\nabla^2_{\boldsymbol\rho}-\frac{1}{2\mu_2}\nabla^2_{\boldsymbol\lambda}+
\frac{e_1e_2}{|{\boldsymbol\rho}|}+\frac{e_1e_3}{|{\boldsymbol\lambda}+\frac{m_2}{m_{12}}{\boldsymbol\rho}|}+
\frac{e_2e_3}{|{\boldsymbol\lambda}-\frac{m_1}{m_{12}}{\boldsymbol\rho}|},
\label{eq3}
\end{equation}
where $m_{12}=m_1+m_2$, $\mu_1=\frac{m_1m_2}{m_1+m_2}$, $\mu_2=\frac{(m_1+m_2)m_3}{m_1+m_2+m_3}$,
$e_1$, $e_2$, $e_3$ are the charges of the helium nucleus, muon, and electron.
It contains the sum of pairwise Coulomb interactions of charged particles. Further refinement \eqref{eq3} 
involves taking into account a number of corrections from the Breit pairwise Hamiltonians: relativistic 
corrections, vacuum polarization effects, and contact interactions. An estimate of the contribution 
of the muon spin-orbit interaction is also given.

For arbitrary states of a muon and an electron with orbital angular momenta $l_1$ and $l_2$, 
a convenient basis for expanding functions that depend on angles is the bipolar spherical harmonics \cite{var}:
\begin{equation}
[Y_{l_1}(\theta_\rho,\phi_\rho)\otimes Y_{l_2}(\theta_\lambda,\phi_\lambda)]_{LM}=
\sum_{m_1,m_2}C^{LM}_{l_1m_1l_2m_2}Y_{l_1m_1}(\theta_\rho,\phi_\rho)Y_{l_2m_2}(\theta_\lambda,\phi_\lambda),
\label{eq4a}
\end{equation}
where $\theta_\rho,\phi_\rho$ and $\theta_\lambda,\phi_\lambda$ are the spherical angles that determine 
the direction of the vectors $\boldsymbol\rho$ and $\boldsymbol\lambda$,
$C^{LM}_{l_1m_1l_2m_2}$ are the Clebsch-Gordon coefficients.
Since in muonic helium the $\mu^-$ meson is in an excited state with orbital angular momentum
$l$, and the electron is in the ground state, the variational wave function of the system is chosen as:
\begin{equation}
\Psi_{lm}({\boldsymbol\rho},{\boldsymbol\lambda},A)=\sum_{i=1}^K C_i 
Y_{lm}(\theta_\rho,\phi_\rho)\rho^l
e^{-\frac{1}{2}\left(A_{11}^i{\boldsymbol\rho}^2+
2A_{12}^i{\boldsymbol\rho}{\boldsymbol\lambda}+A_{22}^i{\boldsymbol\lambda}^2\right)},
\label{eq4}
\end{equation}
where the spherical function $Y_{lm}(\theta_\rho,\phi_\rho)$ describes the angular part of the muon's 
orbital motion. Within the variational approach, solving the Schroedinger equation reduces to solving 
the following matrix problem for the coefficients $C_i$:
\begin{equation}
H\cdot C=EB\cdot C,
\label{eq5}
\end{equation}
where the matrix elements of the Hamiltonian $H_{ij}$ and the normalization $B_{ij}$ can be calculated 
analytically in the basis of Gaussian wave functions. Thus, the normalization of the wave function 
\eqref{eq4} is determined by the following expression:
\begin{equation}
\langle\Psi|\Psi\rangle=\sum_{i,j=1}^K 
C_i C_j 2^{l+2}\pi^{\frac{3}{2}}
\Gamma\left(l+\frac{3}{2}\right)
\frac{B_{22}^l}{(det B)^{l+\frac{3}{2}}},~~~
B_{kn}=A_{kn}^i+A_{kn}^j,
\label{eq6}
\end{equation}
where $\Gamma\big(l+\frac{3}{2}\bigr)$ is the Euler $\Gamma$-function.

Let's consider other analytical results for the Hamiltonian matrix elements. The kinetic energy operator 
contains two terms. The matrix element of the Laplace operator with respect to ${\boldsymbol\lambda}$ 
is:
\begin{equation}
\langle\Psi|\nabla^2_{\boldsymbol\lambda}|\Psi\rangle=\sum_{i,j=1}^K 
\frac{C_i C_j 2^{l+2}\pi^{\frac{3}{2}}}{\langle\Psi|\Psi\rangle}
\Gamma\left(l+\frac{3}{2}\right)
\frac{B_{22}^{l-1}}{(det B)^{l+\frac{5}{2}}}\times
\label{eq7}
\end{equation}
\begin{displaymath}
\left[
3A_{22}^i(A_{22}^i-B_{22})det B+(2l+3)(A_{22}^iB_{12}-A_{12}^iB_{22})^2\right].
\end{displaymath}

A similar matrix element with the Laplace operator on ${\boldsymbol\rho}$ is also expressed through 
nonlinear variational parameters as follows:
\begin{equation}
\langle\Psi|\nabla^2_{\boldsymbol\rho}|\Psi\rangle=\sum_{i,j=1}^K 
\frac{C_i C_j 2^{l+1}\pi^{\frac{3}{2}}}{\langle\Psi|\Psi\rangle}
\Gamma\left(l+\frac{1}{2}\right)
\frac{B_{22}^{l-1}}{(det B)^{l+\frac{5}{2}}}\times
\label{eq8}
\end{equation}
\begin{displaymath}
\left[
(2l+1)det B(-(2l+3)A_{11}^iB_{22}+3(A_{12}^i)^2+2l A_{12}^iB_{12})+(2l+1)(2l+3)(A_{12}^iB_{12}-A_{11}^iB_{22})^2
\right].
\end{displaymath}

The potential energy operator in the nonrelativistic Hamiltonian consists of pairwise Coulomb interactions 
$U_{ij}$ (i, j=1, 2, 3). The matrix elements of the potential energy (in electron atomic units)
are represented by compact expressions for arbitrary states with muon orbital angular momentum $l$:
\begin{equation}
\langle\Psi|U_{12}|\Psi\rangle=-Z\sum_{i,j=1}^K 
\frac{C_i C_j 2^{l+\frac{3}{2}}\pi^{\frac{3}{2}}}{\langle\Psi|\Psi\rangle}
\Gamma\left(l+1\right)
\frac{B_{22}^{l-1}}{(det B)^{l+1}},
\label{eq9}
\end{equation}
\begin{equation}
\langle\Psi|U_{13}|\Psi\rangle=-Z\sum_{i,j=1}^K 
\frac{C_i C_j 2^{l+\frac{5}{2}}\pi}{\langle\Psi|\Psi\rangle}
\Gamma\left(l+\frac{3}{2}\right)
\frac{B_{22}^{l+\frac{1}{2}}}{(det B)^{l+\frac{3}{2}}}
{_2F_1}\left(\frac{1}{2},l+\frac{3}{2},\frac{3}{2},-\frac{(F_2^{23})^2}{det B}\right)
\label{eq10}
\end{equation}
\begin{equation}
\langle\Psi|U_{23}|\Psi\rangle=\sum_{i,j=1}^K 
\frac{C_i C_j 2^{l+\frac{5}{2}}\pi}{\langle\Psi|\Psi\rangle}
\Gamma\left(l+\frac{3}{2}\right)
\frac{B_{22}^{l+\frac{1}{2}}}{(det B)^{l+\frac{3}{2}}}
{_2F_1}\left(\frac{1}{2},l+\frac{3}{2},\frac{3}{2},-\frac{(F_2^{13})^2}{det B}\right)
\label{eq11}
\end{equation}
\begin{equation}
F_2^{13}=B_{12}+\frac{m_1}{m_{12}}B_{22},~~~F_2^{23}=B_{12}-\frac{m_2}{m_{12}}B_{22},
\label{eq12}
\end{equation}
where ${_2F_1}(\alpha,\beta,x)$ is the hypergeometric function.

For $l=1$, expressions \eqref{eq9}-\eqref{eq11} coincide with the previously obtained results \cite{phys2023}. 
Using the matrix elements of the Hamiltonian $\hat H_0$, some energy levels of muonic helium $(\mu^-e^-He)$ are calculated in the Matlab system. The calculations are performed using our program, which was previously 
used to calculate the energy levels of three- and four-particle atoms in quantum electrodynamics 
\cite{apm2026,apm2024,apm2024a}. The calculation results are shown in Table~\ref{tb1}.

It is useful to estimate the contribution of most important corrections to the energy spectrum in order 
to improve the accuracy of calculations.
To this end, we calculate the contribution of several important corrections to the Hamiltonian $\hat H_0$.
The pairwise electromagnetic interaction between particles in quantum electrodynamics is determined 
by the Breit potential \cite{t4}. Among various terms of this potential, we highlight those terms with the 
greatest numerical value. These include relativistic corrections, contact interactions, and corrections 
for vacuum polarization.
The contribution of these corrections is taken into account in the first order of perturbation theory.

Relativistic corrections are defined in the energy spectrum by the following terms in electron atomic units 
(e.a.u.):
\begin{equation}
\Delta U_{rel}=-\frac{\alpha^2}{8}\left(\frac{{\bf p}_1^4}{m_1^3}+\frac{{\bf p}_2^4}{m_2^3}+
\frac{{\bf p}_3^4}{m_3^3}\right).
\label{eq13}
\end{equation}

The main contribution to the relativistic correction \eqref{eq13} is connected with the electron's motion. 
The value of corresponding matrix element from $\Delta U^e_{rel}$ can be obtained in exactly 
the same way as \eqref{eq7} in terms of variational parameters:
\begin{equation}
\langle\Psi|-\frac{\alpha^2}{8}\nabla^4_{\boldsymbol\lambda}|\Psi\rangle=-\alpha^2\sum_{i,j=1}^K 
\frac{C_i C_j 2^{l-1}\pi^{\frac{3}{2}}}{\langle\Psi|\Psi\rangle}\Gamma\left(l+\frac{3}{2}\right)
\frac{B_{22}^{l-2}}{(det B)^{l+\frac{7}{2}}}\Bigl[15(A_{22}^i)^2 (det B)^2(A_{22}^i-B_{22})^2+
\label{eq14}
\end{equation}
\begin{displaymath}
10 (2l+3)  A_{22}^i(A_{22}^i-B_{22})det B(A_{22}^iB_{12}-A_{12}^i B_{22})^2+
(2l+3)(2l+5)(A_{22}^iB_{12}-A_{12}^iB_{22})^4\Bigr].
\end{displaymath}

\begin{table}[htbp]
\caption{\label{tb1} Energy levels of the high-angular-momentum muon states 
of muonic helium obtained in the nonrelativistic 
approximation and the values of the main corrections in the energy spectrum in electron atomic units (e.a.u.).}
\bigskip
\begin{tabular}{|c|c|c|c|c|}   \hline
State&  $E_{nr}$    & $-\frac{\alpha^2}{8}{\bf p}_e^4$  & $\Delta U_{vp}$ & $\Delta U_{cont}$ \\
   &    &    &     &          \\             \hline
  \multicolumn{5}{|c|}{$({_2^3}He-\mu^-- e^-)$ atom}     \\            \hline
 (14,13)&     -2.81111089   &-0.00030330 & -0.00000033   & 0.00000361   \\     \hline
 (14,12)&     -2.84289819  & -0.00031998 & -0.00000035   & 0.00000380    \\    \hline
 (13,12)&     -3.07090707  & -0.00025213 & -0.00000028   & 0.00000303   \\     \hline
 (13,11)&     -3.10157048  & -0.00027561 & -0.00000030   & 0.00000324     \\    \hline
 (12,11)&     -3.42617315  & -0.00020518 & -0.00000024   & 0.00000247   \\     \hline
 (12,10)&     -3.45844905  & -0.00019336 & -0.00000023   & 0.00000233     \\    \hline
   \multicolumn{5}{|c|}{$({_2^4}He-\mu^-- e^-)$ atom}     \\            \hline    
 (14,13)&    -2.82505929 & -0.00030182      & -0.00000033  & 0.00000361 \\     \hline
 (14,12)&    -2.85670449 & -0.00031536      & -0.00000034  & 0.00000370  \\    \hline
 (13,12)&    -3.08873955 & -0.00019461      & -0.00000023  & 0.00000245 \\     \hline
 (13,11)&    -3.11916631  &-0.00027631      & -0.00000029  & 0.00000309  \\    \hline
 (12,11)&    -3.44854579  &-0.00019806      & -0.00000024  & 0.00000242  \\     \hline
 (12,10)&    -3.48181792  &-0.00021110      & -0.00000025  & 0.00000235  \\    \hline
\end{tabular}
\end{table}

The Breit potential also includes a relativistic recoil correction; however, since the electron mass 
is much smaller than the masses of the muon and the nucleus, we neglect it at this stage of the calculations.

We also take into account the effects of vacuum polarization in the energy spectrum \cite{apm2022}. 
Since the Compton wavelength 
of both the electron and the muon in a highly excited state is much smaller than the radius of the Bohr orbit, 
we can use the following expression for the vacuum polarization potential in e.a.u. 
(the $\delta$-approximation for the Uehling-Serber potential) \cite{egs}:
\begin{equation}
\Delta U_{vp}=\Delta U_{vp}(r_{13})+\Delta U_{vp}(r_{23})=
-\frac{4}{15}\alpha^2(Z\alpha)\delta({\boldsymbol\lambda}+\frac{m_2}{m_{12}}{\boldsymbol\rho})
+\frac{4}{15}\alpha^2(Z\alpha)\delta({\boldsymbol\lambda}-\frac{m_1}{m_{12}}{\boldsymbol\rho}).
\label{eq15}
\end{equation}

The matrix elements of such potentials are calculated analytically using \eqref{eq2} in closed form:
\begin{equation}
<\Psi|\Delta U_{vp}(r_{13})|\Psi>=-\frac{4}{15}\alpha^2(Z\alpha)\sum_{i,j=1}^K 
\frac{C_i C_j 2^{l+\frac{1}{2}}}{\langle\Psi|\Psi\rangle}
\Gamma\left(l+\frac{3}{2}\right)\frac{1}{\left(F_1^{13}\right)^{l+\frac{3}{2}}},
\label{eq16}
\end{equation}
\begin{equation}
<\Psi|\Delta U_{vp}(r_{23})|\Psi>=-\frac{4}{15}\alpha^2(Z\alpha)\sum_{i,j=1}^K 
\frac{C_i C_j 2^{l+\frac{1}{2}}}{\langle\Psi|\Psi\rangle}
\Gamma\left(l+\frac{3}{2}\right)\frac{1}{\left(F_1^{23}\right)^{l+\frac{3}{2}}},
\label{eq17}
\end{equation}
\begin{equation}
F_1^{13}=B_{11}+\frac{m_2^2}{m_{12}^2}B_{22}-2\frac{m_2}{m_{12}}B_{12},~~~
F_1^{23}=B_{11}+\frac{m_1^2}{m_{12}^2}B_{22}+2\frac{m_1}{m_{12}}B_{12}.
\label{eq18}
\end{equation}

The contact interaction potential, like \eqref{eq15}, is expressed through $\delta$-functions 
in the form (in electron atomic units):
\begin{equation}
\Delta U_{cont}=
\frac{\pi Z\alpha^2}{2}\delta({\boldsymbol\lambda}+\frac{m_2}{m_{12}}{\boldsymbol\rho})
-\frac{\pi \alpha^2}{2}\delta({\boldsymbol\lambda}-\frac{m_1}{m_{12}}{\boldsymbol\rho}).
\label{eq18a}
\end{equation}

Table~\ref{tb1} presents the results of calculating the energy values using the Hamiltonian
$\hat H_0$ and the values of the matrix elements \eqref{eq14}, \eqref{eq16}, \eqref{eq17}, \eqref{eq18a}.
\begin{figure}[htbp]
\centering
\includegraphics[scale=0.7]{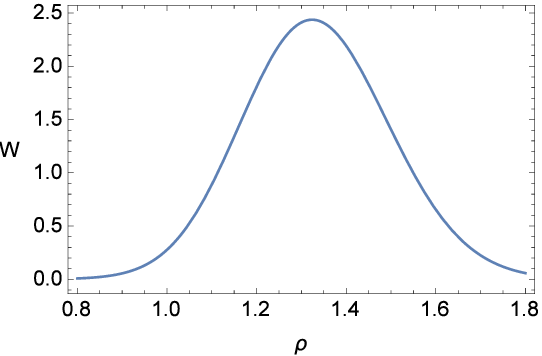}
\includegraphics[scale=0.7]{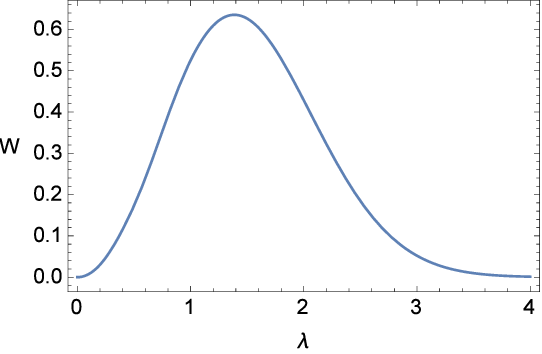}
\caption{Radial distribution densities $W(\rho)$, $W(\lambda)$ for the $(14,13)$ state of 
$({^4_2He}-\mu^--e^-)$.
The values of the variables $\rho$ and $\lambda$ are expressed in electron atomic units.}
\label{pic1}
\end{figure}
\begin{figure}[htbp]
\centering
\includegraphics[scale=0.7]{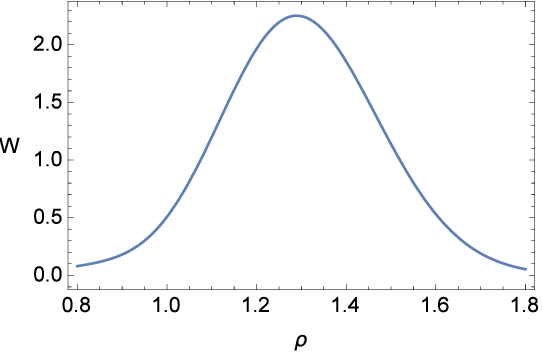}
\includegraphics[scale=0.7]{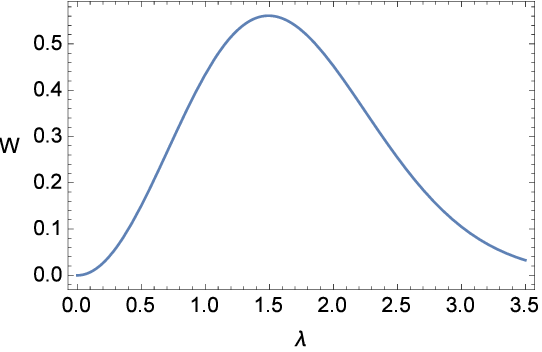}
\caption{Radial distribution densities $W(\rho)$, $W(\lambda)$ for the $(13,12)$ state of $({^4_2He}-\mu^--e^-)$.
The values of the variables $\rho$ and $\lambda$ are expressed in electron atomic units.}
\label{pic2}
\end{figure}

An estimate of possible value of the spin-orbit correction for the muon in the $(14,13)$-state 
is presented in Appendix~\ref{app2}.

To study the internal structure of three-particle bound states, we calculate the mean-square distances 
between particles. Corresponding matrix elements are calculated analytically, and the results are presented 
in Appendix~\ref{app1}. Using the relations \eqref{a1}-\eqref{a3}, we present numerical values 
of the mean-square distances between particles in the $(^4He-\mu^--e^-)$ atom for the $(14,13)$-state, 
numbering the particles as follows: $He\to 1$, $\mu^-\to 2$, $e^-\to 3$:
\begin{equation}
\label{f19}
\delta_{12}=\sqrt{\langle({\bf r}_1-{\bf r}_2)^2\rangle}=1.34~(0.71\times 10^{-8}~cm),
\end{equation}
\begin{equation}
\label{f20}
\delta_{13}=\sqrt{\langle({\bf r}_1-{\bf r}_3)^2\rangle}=0.47~(0.25\times 10^{-8}~cm),~
\end{equation}
\begin{equation}
\label{f21}
\delta_{23}=\sqrt{\langle({\bf r}_2-{\bf r}_3)^2\rangle}=1.08~(0.57\times 10^{-8}~cm).
\end{equation}

Average distances between particles are presented in \eqref{f19}-\eqref{f21} in electron atomic units 
and in cm (in parentheses).

Some details of the internal structure of bound states can also be obtained by constructing the radial 
distribution densities for $\rho$ and $\lambda$ and calculating the root-mean-square values 
of $\langle\rho^2\rangle$,
$\langle\lambda^2\rangle$, which are determined by the expressions:
\begin{equation}
\label{eq19}
W(\rho)=\frac{(2\pi)^{\frac{3}{2}}}{\langle\Psi|\Psi\rangle}\sum_{i,j=1}^K \frac{C_i C_j}{B_{22}^{3/2}}
\rho^{(2l+2)}e^{-\frac{1}{2}\frac{det B}{B_{22}}\rho^2},
\end{equation}
\begin{equation}
\label{eq20}
W(\lambda)=\frac{2^{l+\frac{5}{2}}\pi}{\langle\Psi|\Psi\rangle}\sum_{i,j=1}^K \frac{C_i C_j
\Gamma\left(l+\frac{3}{2}\right)}
{B_{11}^{l+\frac{3}{2}}}\lambda^2
e^{-\frac{1}{2}B_{22}\lambda^2}{_1F_1}\left(l+\frac{3}{2},\frac{3}{2},\frac{B_{12}^2\lambda^2}{2B_{11}}\right),
\end{equation}
\begin{equation}
\label{eq21}
W(\rho,\lambda)=\frac{4\pi}{\langle\Psi|\Psi\rangle}\sum_{i,j=1}^K \frac{C_i C_j}{B_{12}}\rho^{2l+1}\lambda
e^{-\frac{1}{2}[B_{11}\rho^2+B_{22}\lambda^2]}sh(B_{12}\rho\lambda),~ B_{lk}=A^i_{lk}+A^j_{lk},
\end{equation}
\begin{equation}
\label{eq22}
\langle\rho^2\rangle=
\frac{\pi^{\frac{3}{2}}2^{l+3}\Gamma\left(l+\frac{5}{2}\right)}{\langle\Psi|\Psi\rangle}
\sum_{i,j=1}^K C_i C_j\frac{B_{22}^{l+1}}
{(det B)^{l+5/2}},
\end{equation}
\begin{equation}
\label{eq23}
\langle\lambda^2\rangle=
\frac{\pi^{\frac{3}{2}}2^{l+2}\Gamma\left(l+\frac{3}{2}\right)}{\langle\Psi|\Psi\rangle}\sum_{i,j=1}^K C_i C_j
\frac{B_{22}^{l-1}}{(det B)^{l+\frac{5}{2}}} (3B_{11}B_{22}+2B_{12}^2l).
\end{equation}

\begin{figure}[htbp]
\centering
\includegraphics[scale=0.4]{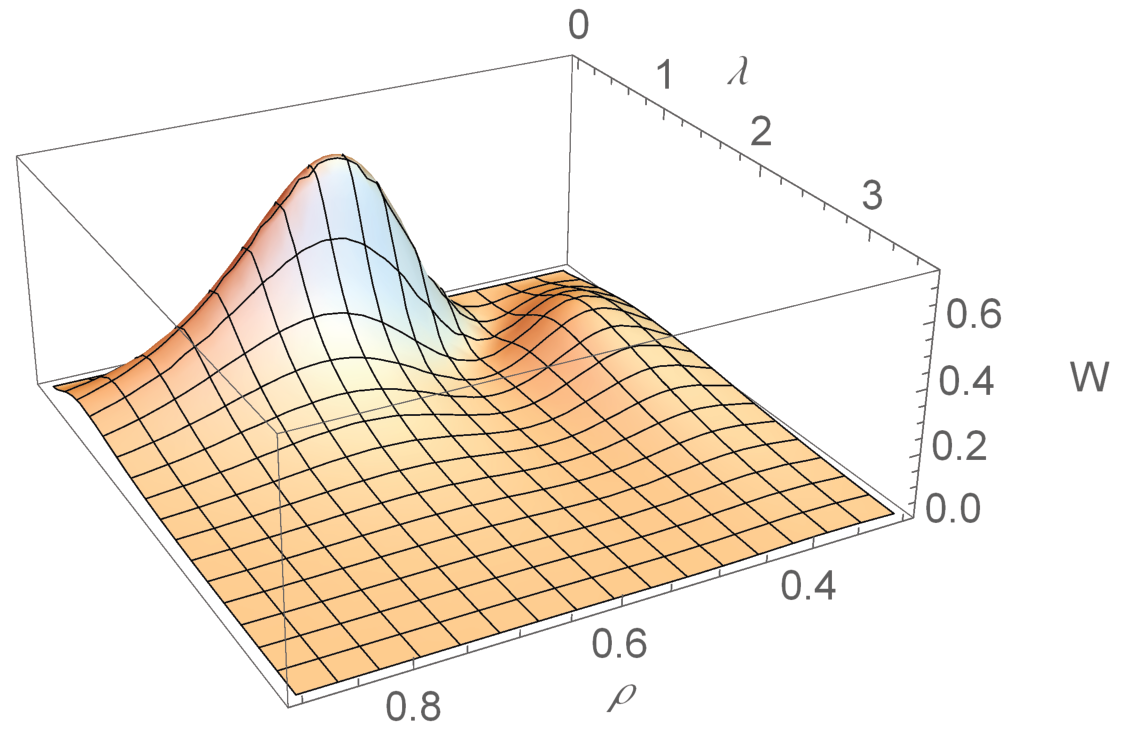}
\includegraphics[scale=0.4]{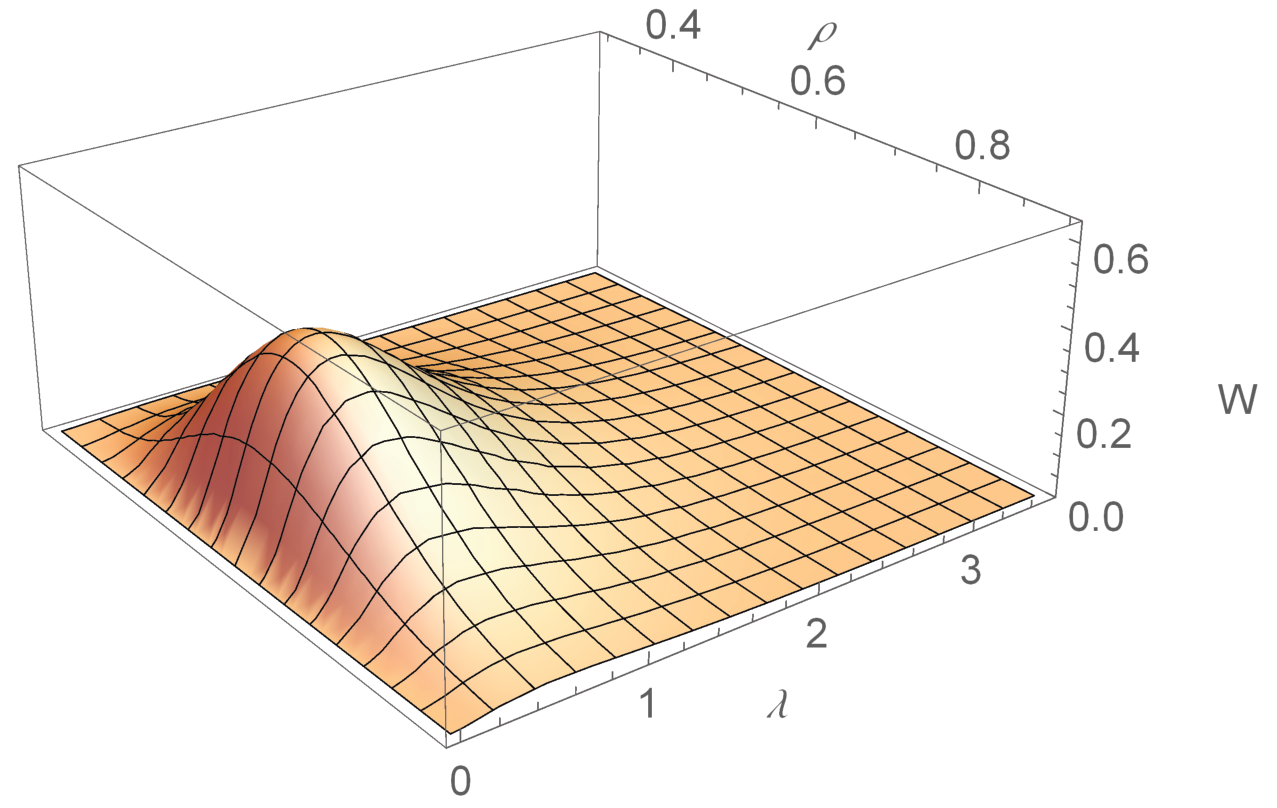}
\caption{Radial distribution density $W(\rho,\lambda)$ for the states (14,13) $(\mu^-- e^-- ^4_2He)$ (right)
and (14,12) $(\mu^-- e^-- ^3_2He)$ (left).
The values of the variables $\rho$ and $\lambda$ are expressed in electron atomic units.
}
\label{pic5}
\end{figure}

The radial distribution densities are shown in Figs.~\ref{pic1} and 
\ref{pic2} for muonic helium-4. These graphs demonstrate 
the presence of characteristic distances in three-particle systems
$(^{3,4}He-\mu^--e^-)$. From the graphs in Figs.~\ref{pic1} and 
\ref{pic2}, it follows that for the states under consideration,
the muon is located at approximately the same distances from the nucleus as the electron.
The distribution densities for the two radial variables, rho and lambda, provide a more complete picture 
of the characteristic distances in this three-particle system. They are shown in Fig.~\ref{pic5}.

Note that we do not consider the fine splitting in a three-particle atom, determined by the interaction 
of the electron spin and the large orbital angular momentum of the muon.

\section{Conclusion}

A field of research in physics has recently emerged devoted to studying the properties of exotic atoms
\cite{ncm,scheck,kaon}. 
These include a wide variety of systems. These atoms have short lifetimes, but they also have a relatively 
simple structure (two-particle and three-particle), allowing for the use in their study of rigorous methods and 
the calculation of observable properties with a high degree of accuracy. Experiments conducted 
with these atoms allow us to take a look into the field of particle interactions inaccessible to conventional systems.
Spectroscopy of various exotic atoms and molecules can provide new information about the nature of fundamental
interactions and the values of fundamental parameters of the Standard Model. 
For example,the PiHe collaboration performed laser spectroscopy of the infrared transition 
in three-particle pion helium atoms \cite{hori1,hori2}. These atoms were created in a superfluid (He-II) 
helium target. Similar measurements in antiproton helium atoms embedded in liquid helium were performed 
by the CERN ASACUSA collaboration \cite{asacusa}. As a result of such experiments, it is possible 
to determine a more precise ratio of the masses of the antiproton and electron, and the pion and electron.
A similar program for studying kaonic helium is planned for a more accurate measurement of the K-meson mass.

The exotic atom of muonic helium ($\mu^-- e^-- He$) is formed by the capture of a negative muon by the helium atom 
as a result of electromagnetic interaction. It should be noted that research into muonic, pionic, and kaonic 
helium, as well as the various processes connected with their formation, has been underway for quite some 
time. Initially, a beam of muons (pions, kaons) slows down in matter. Once a muon reaches a state 
in which its kinetic energy is zero, it is captured by an atom and transitions to states with high 
orbital angular momentum, forming a muonic atom. The exact distribution of these states is unknown 
experimentally. Since all low-lying muonic states are unoccupied, the muon quickly transitions 
to the lowest accessible quantum state (1S). Auger and radiative transitions are responsible 
for this de-excitation. The study of muonic helium was fraught with various technical and scientific 
difficulties, including the availability of intense muon beams, accurate X-ray spectroscopy, 
and effective analysis of the background radiation. It can be said that these difficulties 
have been largely overcome. The measured energies of the (n,l)-(1s) transitions in the case 
of $(\mu^--He^{++})$ with $n\leq 6$ have been obtained in many studies and are in good agreement 
with theoretical estimates. The emitted radiation for transitions to the lowest-lying levels 
is in the X-ray energy range.

In this work, we investigate excited states in neutral muonic helium, a system that exhibits similarities 
to the muonic helium ion but is clearly distinct from it \cite{cohen,landua,baird,wetmore,souder,egger}. 
Therefore, a detailed comparison of our numerical 
results with previous calculations for neutral muonic helium proved difficult due to the virtual absence 
of such calculations. We calculate the energy levels of high-lying orbits that overlap significantly 
with the electron shells of the atom. These orbits are characterized by a principal quantum number $n\sim 14$. 
The studied principal quantum number distributions in \cite{cohen} 
confirm this well. Measurements of the 
X-ray intensity of such high-level transitions (as well as Auger transitions), when compared with muon 
cascade calculations, provide information on the initial population of muonic states with quantum numbers 
$(n,l)$. In this way, information is obtained regarding the probabilities of atomic muon capture, 
which is determined by the fraction of particles that are localized in helium atoms by replacing 
valence electrons.

We study the energy levels of muonic helium for states in which the muon has such a large orbital 
angular momentum that it is approximately at the same distance from the nucleus as the electron. Calculations are performed in leading order using the variational method with a Gaussian basis, and several key corrections 
determined by the Breit Hamiltonian (relativism, vacuum polarizations
and contact interactions) are calculated in first-order perturbation theory. Since the electron is in the $1S$ 
state, the three-particle state in Table ~\ref{tb1} is denoted by $(n,l)$, where $l$ is the muon's orbital 
angular momentum and $n$ is the principal quantum number of the $(\mu^- He^{++})$ subsystem.
When calculating the energy of a specific state, we perform an initial run of the program 10 times with 
a minimum basis of K=200 functions, varying the interval boundaries to generate nonlinear variational 
parameters. Having chosen these boundaries so that the minimum energy is achieved at this stage, 
we run several calculation programs with the same parameters, increasing the basis in the \eqref{eq4} 
expansion to approximately K=800 functions. At each step of the program, the basis is incremented 
by one, and refinement cycles are performed. The number of such cycles (usually 1000) is specified 
in the input file at the beginning of the program.
The numerical results in the Table~\ref{tb1} (the main contribution and corrections) are presented with 
an accuracy of 8 decimal places, which remains unchanged for our basis sizes. Previous studies using 
a basis of Gaussian wave functions show that it yields results consistent with previous results obtained 
using other approaches \cite{apm2026,apm2024,apm2024a}.

Our previous studies of the energy levels of both pionic and kaonic helium were performed within 
the  variational approach developed in \cite{apm2024,apm2024a}. An important addition to our previous 
studies is the calculation of the energy levels of excited states of muonic helium, 
which is performed in this paper. 
The numerical results obtained in Table~\ref{tb1} can be used in parallel to analyze experimental data 
for both pionic helium and muonic helium, and measuring the transition frequencies between the 
high-angular-momentum muon states in muonic helium opens a new opportunity for determining the muon mass.
A direct comparison of our results with experimental data is impossible due to the lack of experimental data. 
However, the obtained energy values are important for energy analysis of the complex processes that occur 
in the helium atom after the replacement of one electron by a muon.

\begin{acknowledgments}
The authors are grateful to V. I. Korobov for helpful discussions.
This work was supported by the Russian Science Foundation (grant no. RSF 25-72-00029) (F.A.M.).
\end{acknowledgments}

\appendix
\section{Root mean square distances between particles.}
\label{app1} 

To calculate the mean-square distances between particles, we use the explicit form of the wave function 
of the system \eqref{eq2}. The results of the analytical integration are as follows:
\begin{equation}
\label{a1}
\langle({\bf r}_1-{\bf r}_2)^2\rangle=\sum_{i,j=1}^K\frac{C_i C_j}{\langle\Psi|\Psi\rangle} 
2^{(l+3)}\pi^{\frac{3}{2}}\Gamma(l+\frac{5}{2})\frac{B_{22}^{(l+1)}}{(\det B)^{(l+\frac{5}{2})}},
\end{equation}
\begin{equation}
\label{a2}
\langle({\bf r}_1-{\bf r}_2)^2\rangle=\sum_{i,j=1}^K\frac{C_i C_j}{\langle\Psi|\Psi\rangle} 
2^{(l+2)}\pi^{\frac{3}{2}}\Gamma(l+\frac{3}{2})\frac{B_{22}^{(l-1)}}{(\det B)^{(l+\frac{5}{2})}}
\Bigl[(2l+3)\bigl(B_{12}-\frac{m_2}{m_{12}}B_{22}\bigr)^2+3\det B\Bigr],
\end{equation}
\begin{equation}
\label{a3}
\langle({\bf r}_2-{\bf r}_3)^2\rangle=\sum_{i,j=1}^K\frac{C_i C_j}{\langle\Psi|\Psi\rangle} 
2^{(l+2)}\pi^{\frac{3}{2}}\Gamma(l+\frac{3}{2})\frac{B_{22}^{(l-1)}}{(\det B)^{(l+\frac{5}{2})}}
\Bigl[(2l+3)\bigl(B_{12}+\frac{m_1}{m_{12}}B_{22}\bigr)^2+3\det B\Bigr].
\end{equation}

Using the general formulas \eqref{a1}-\eqref{a3} one can obtain numerical values of the average 
distances between particles.

\section{Spin-orbit correction in energy level splitting}
\label{app2} 

The Breit interaction potential contains a correction for the spin-orbit interaction of the muon, 
since muon states with large values of orbital angular momentum are considered.
Due to the presence of three particles, the spin-orbit interaction contains several terms.
Spin-orbit interactions are typically divided into self and mutual spin-orbit interactions 
\cite{araki,sucher,korobov2022}.
Here we want to give only a preliminary numerical estimate of some of the main terms.

One of them is the spin-orbit interaction of the muon, which is determined by the following term in the Hamiltonian:
\begin{equation}
\label{b1}
\Delta V^\mu_{LS}(\rho)=\frac{Z\alpha}{\rho^3}\frac{({\bf s}_\mu{\bf L})}{2m_\mu^2},
\end{equation}
where ${\bf L}={\bf L}_{12}$ is the muon orbital momentum.

The general matrix element of the operator \eqref{b1} contains integration over the coordinate $\rho$, 
which gives the following result:
\begin{equation}
\label{b2}
\langle\frac{1}{\rho^3}\rangle=\frac{2^{l+1/2}\pi^{3/2}\Gamma(l)B_{22}^{l-3/2}}{(\det B)^l}.
\end{equation}

In general, the splitting formula between energy levels with muon total momentum $j_\mu=l+\frac{1}{2}$ 
and $j_\mu=l-\frac{1}{2}$ (${\bf J}_\mu={\bf L}+{\bf s}_\mu$) then takes the form 
(in electron atomic units, $\tilde m_\mu=m_\mu/m_e$):
\begin{equation}
\label{b3}
\Delta E^\mu_{LS}=\langle \Delta V^\mu_{LS}\rangle(j_\mu=l+\frac{1}{2})-\langle \Delta V^\mu_{LS}
\rangle(j_\mu=l-\frac{1}{2})=
\frac{Z\alpha^2(2l+1)}{4\tilde m_\mu^2}\sum_{i,j=1}^K \frac{C_i C_j}{\langle\Psi|\Psi\rangle}
\frac{2^{l+1/2}\pi^{3/2}\Gamma(l)B_{22}^{l-3/2}}{(det B)^{l}}.
\end{equation}

To numerically estimate the splitting using this formula, we consider the state (14,13) with 
the orbital angular momentum of the muon $l=13$. Using the obtained values of the variational 
parameters, we find the following splitting value:
\begin{equation}
\label{b4}
\Delta E^\mu_{LS}\left(j_\mu=\frac{27}{2},j_\mu=\frac{25}{2}\right)=0.00000011~e.a.u=3\times 10^{-6}~eV.
\end{equation}

This value is close to the correction for vacuum polarization (see Table~\ref{tb1})
and at this stage such corrections can be neglected.

A more significant contribution is made by the spin-orbit interaction of the electron, 
since in the denominator the muon mass is replaced by the electron mass.
The corresponding spin-orbit interaction operator contains the relative distance between 
the muon and the electron ${\bf r}_{23}={\boldsymbol\lambda}-\frac{m_1}{m_{12}}{\boldsymbol\rho}$:
\begin{equation}
\label{b5}
\Delta V^e_{LS}(r_{23})=-\frac{\alpha}{r_{23}^3}\frac{({\bf s}_e{\bf L})}{2m_e^2},
\end{equation}
where relative electron-muon orbital momentum ${\bf L}_{23}={\bf L}_{12}={\bf L}$.

When calculating the matrix element \eqref{b5} using the wave functions \eqref{eq5}, the integration variable is 
shifted $({\boldsymbol\lambda}\to{\boldsymbol\lambda}+\frac{m_1}{m_{12}}{\boldsymbol\rho})$, 
and integration with respect to $\rho$  is performed analytically. For further integration 
with respect to ${\boldsymbol\lambda}$, the following formula is used \cite{araki,apm2025}:
\begin{equation}
\label{b6}
\langle\frac{1}{4\pi\lambda^3}\rangle=\lim_{\varepsilon\to 0}\langle\frac{\Theta(\lambda-\varepsilon)}{4\pi\lambda^3}+
\Bigr(\ln\varepsilon+\frac{C}{2}\Bigl)\delta({\boldsymbol\lambda})\rangle,
\end{equation}
where $C$ is the Euler constant. Formula \eqref{b6} is used for states with different values of orbital 
angular momentum $l$. Thus, in the case of state $(14,13)$, the splitting of the energy level due to 
the spin-orbit interaction of the electron is determined by the following expression 
(${\bf J}_e={\bf L}+{\bf s}_e$):
\begin{equation}
\label{b7}
\Delta E^e_{LS}\left(j_e=\frac{27}{2},j_e=\frac{25}{2}\right)=
-\frac{27\alpha^2}{4}\sum_{i,j=1}^K \frac{C_i C_j}{\langle\Psi|\Psi\rangle}
\frac{2^{\frac{31}{2}}\Gamma(\frac{29}{2})\pi}{D^\frac{29}{2}}\Bigl[\frac{2^{22}}{456326325}\omega^{13}+
\end{equation}
\begin{displaymath}
\frac{2^{19}}{3900225}\omega^{12}+\frac{2^{19}}{572033}\omega^{11}+\frac{2^{17}}{33915}\omega^{10}+
\frac{2^{15}}{2907}\omega^{9}+\frac{2^{11}}{85}\omega^{8}+\frac{2^{12}}{105}\omega^{7}+
\frac{2^{10}}{21}\omega^{6}+
\end{displaymath}
\begin{displaymath}
\frac{1664}{35}\omega^{5}+
\frac{2288}{63}\omega^{4}+
\frac{2288}{105}\omega^{3}+\frac{52}{5}\omega^{2}+\frac{13}{3}\omega\Biggr],
\end{displaymath}
\begin{equation}
\label{b8}
\omega=\frac{(B_{12}+\frac{m_1}{m_{12}}B_{22})^2}{\det B},~~~D=B_{11}+2\frac{m_1}{m_{12}}B_{12}+
\frac{m_1^2}{m_{12}^2}B_{22}.
\end{equation}

The numerical value of the splitting \eqref{b7} is equal to
\begin{equation}
\label{b9}
\Delta E^e_{LS}\left(j_e=\frac{27}{2},j_e=\frac{25}{2}\right)=-0.00040050~e.a.u=0.010898~eV.
\end{equation}

This estimate demonstrates that, in a precision study of energy levels, spin-orbit interaction 
must be considered along with other corrections from the Table~\ref{tb1}. This requires taking into account 
not only the spin-orbit interaction but also the spin-spin interaction with the account of recoil effects. 
A more detailed study of fine and hyperfine structure effects in the case of excited states in
muonic helium is planned for future studies.


\begin{thebibliography}{99}

\bibitem{H1}B.~de~Beauvoir, F.~Nez, L.~Julien, et al.,
Absolute Frequency Measurement of the 2S-8S/D Transitions in Hydrogen and Deuterium: New Determination
of the Rydberg Constant, Phys. Rev. Lett. {\bf 78}, 440 (1997);
https://doi.org/10.1103/PhysRevLett.78.440.

\bibitem{H2}C.~Schwob, L.~Jozefowski, B.~de~Beauvoir, et al.,
Optical Frequency Measurement of the 2S-12D Transitions in Hydrogen and Deuterium: Rydberg
Constant and Lamb Shift Determinations, Phys. Rev. Lett. {\bf 82}, 4960 (1999);
https://doi.org/10.1103/PhysRevLett.82.4960.

\bibitem{RA}T.~F.~Gallagher, Rydberg atoms, Cambridge University Press, NY, 1994.

\bibitem{merkt}S.~Scheidegger and F.~Merkt, Precision-spectroscopic determination of the binding
energy of a two-body quantum system: The hydrogen atom and the proton-size
puzzle, Phys. Rev. Lett. {\bf 132}, 113001 (2024);
https://doi.org/10.1103/PhysRevLett.132.113001.

\bibitem{aldo}A.~Antognini, F.~Hagelstein, and V.~Pascalutsa, The proton structure in and out 
of muonic hydrogen, Ann. Rev. Nucl. Part. Sci. {\bf 72}, 389 (2022);
https://doi.org/10.1146/annurev-nucl-101920-024709.

\bibitem{uj}U.~D.~Jentschura and D.~C.~Yost,
Precision Rydberg state spectroscopy with slow electrons
and the proton-radius puzzle,
Phys. Rev. A {\bf 108}, 062822 (2023);  https://doi.org/10.1103/PhysRevA.108.062822.

\bibitem{egs}M.~I.~Eides, H.~Grotch and V.~A.~Shelyuto, Theory of light hydrogenlike atoms,
Phys. Rept. {\bf 342}, 63 (2001);
https://doi.org/10.1016/S0370-1573(00)00077-6.

\bibitem{hori1}M.~Hori, A.~Soter, and V.~I.~Korobov, 
Proposed method for laser spectroscopy of pionic helium atoms to determine the charged-pion mass,
Phys. Rev. A {\bf 89}, 042515 (2014);
https://doi.org/10.1103/PhysRevA.89.042515.

\bibitem{hori2}M.~Hori, H.~Aghai-Khozani,  A.~S\'oter, A.~Dax, and D.~Barna, 
Laser spectroscopy of pionic helium atoms,
Nature {\bf 581}, 37 (2020);
https://doi.org/10.1038/s41586-020-2240-x.

\bibitem{bakalov}D.~Bakalov and B.~Obreshkov,
Collisional shift and broadening of the transition lines in pionic helium,
Phys. Rev. {\bf 93}, 062505 (2016);
https://doi.org/10.1103/PhysRevA.93.062505.

\bibitem{crystal}M.~Trassinelli, D.~F.~Anagnostopoulos, G.~Borchert et al.,
Measurement of the charged pion mass using X-ray spectroscopy of exotic atoms,
Phys. Lett. B {\bf 759}, 583 (2016);
https://doi.org/10.1016/j.physletb.2016.06.025.

\bibitem{pdg}S.~Navas et al. (Particle Data Group), Review of Particle Physics, 
Phys. Rev. D {\bf 110}, 030001 (2024);
https://doi.org/10.1103/PhysRevD.110.030001.

\bibitem{phys2023}A.~V.~Eskin, V.~I.~Korobov, A.~P.~Martynenko and F.~A.~Martynenko,
Energy Levels of Three-Particle Muon Electron Helium in Variational Approach,
Phys. Atom. Nucl. {\bf 86}, 4, 583 (2023);
https://doi.org/10.1134/S106377882304021X.

\bibitem{apm2026}A.~V.~Eskin,  A.~P.~Martynenko, F.~A.~ Martynenko, and D.~K.~Pometko,
Few-particle lepton bound states in variational approach,
Int. J. Mod. Phys. A {\bf 41}, no.15n16, 2650108 (2026);
https://doi.org/10.1142/S0217751X26501083.

\bibitem{apm2024}V.~I.~Korobov, F.~A.~Martynenko, A.~P.~Martynenko, and A.~V.~Eskin,
Energy Levels of Pionic and Kaonic Helium in the Variational Approach,
Phys. Part. Nucl. {\bf 55}, 4, 705 (2024);
https://doi.org/10.1134/S1063779624700047.

\bibitem{apm2024a}V.~I.~Korobov, A.~V.~Eskin, A.~P.~Martynenko, and F.~A.~Martynenko,
Energy levels of mesonic helium in quantum electrodynamics,
Phys. Rev. A {\bf 109}, 3, 032802 (2024);
https://doi.org/10.1103/PhysRevA.109.032802.

\bibitem{mohr}S.~D.~Lakdawala and P.~Mohr, Hyperfine structure in muonic helium,
Phys. Rev. A {\bf 22}, 1572 (1980);
https://doi.org/10.1103/PhysRevA.22.1572.

\bibitem{huang}K.~N.~Huang and V.~W.~Hughes, 
Theoretical hyperfine structure of the muonic $^3He$ and $^4He$ atoms,
Phys. Rev. A {\bf 26}, 2330 (1982);
https://doi.org/10.1103/PhysRevA.26.2330.

\bibitem{amusia}M.~Ya.~Amusia, M.~Ju.~Kuchiev, and V.~L.~Yakhontov, 
Computation of the hyperfine structure in the $(\alpha-\mu^--e^-)^0$ atom,
J. Phys. B {\bf 16}, L71 (1983);
https://doi.org/10.1088/0022-3700/16/3/007.

\bibitem{amusia1}S.~G.~Karshenboim, V.~G.~Ivanov, and M.~Ya.~Amusia,
Lamb shift of electronic states in neutral muonic helium, an electron-muon-nucleus system,
Phys. Rev. A {\bf 91}, 3, 032510 (2015);
https://doi.org/10.1103/PhysRevA.91.032510.

\bibitem{apm2008}A.~A.~Krutov and A.~P.~Martynenko,
Ground-state hyperfine structure of the muonic helium atom,
Phys. Rev. A {\bf 78}, 032513 (2008);
https://doi.org/10.1103/PhysRevA.78.032513.

\bibitem{apm2022}R.~N.~Faustov, V.~I.~ Korobov,  A.~P.~Martynenko, and F.~A.~ Martynenko,
Ground-state hyperfine structure of light muon-electron ions,
Phys. Rev. A {\bf 105}, 042816 (2022);
https://doi.org/10.1103/PhysRevA.105.042816.

\bibitem{apm2021}A.~E.~Dorokhov, V.~I.~ Korobov,  A.~P.~Martynenko, and F.~A.~ Martynenko,
Low-lying electron energy levels in three-particle electron-muon ions of Li, Be, and B,
Phys. Rev. A {\bf 103}, 052806 (2021);
https://doi.org/10.1103/PhysRevA.103.052806.

\bibitem{rd}R.~J.~Drachman, 
Nonrelativistic hyperfine splitting in muonic helium by adiabatic perturbation theory,
Phys. Rev. A {\bf 22}, 1755 (1980);
https://doi.org/10.1103/PhysRevA.22.1755.

\bibitem{melezhik}S.~I.~Vinitsky, V.~S.~Melezhik, I.~I.~Ponomarev et al.,
Calculation of Energy Levels of Hydrogen Isotope 
$\mu$ Mesic Molecules in the Adiabatic Representation of Three-body Problem,
Sov. Phys. JETP {\bf 52}, 353 (1980).

\bibitem{puchalski1}M.~Puchalski, D.~Kedziera, and K.~Pachucki,
Ground state of Li and $Be^+$ using explicitly correlated functions,
Phys. Rev. A {\bf 80}, 032521 (2009); http://dx.doi.org/10.1103/PhysRevA.80.032521.

\bibitem{puchalski2}M.~Puchalski and K.~Pachucki,
Applications of four-body exponentially correlated functions,
Phys. Rev. A {\bf 81}, 052505 (2010); http://dx.doi.org/10.1103/PhysRevA.81.052505.

\bibitem{frolov}A.~M.~Frolov, 
Properties and hyperfine structure of helium-muonic atoms,
Phys. Rev. A {\bf 61}, 022509 (2000); https://doi.org/10.1103/PhysRevA.61.022509.

\bibitem{frolov1}A.~M.~Frolov, 
The hyperfine structure of the ground states in the helium-muonic atoms,
Phys. Lett. A {\bf 376}, 2548 (2012);
http://dx.doi.org/10.1016/j.physleta.2012.06.024.

\bibitem{chen}M.-K.~Chen, Correlated wave functions and hyperfine splitting of the 2s state of muonic
$^{3,4}He$ atoms, Phys. Rev. A {\bf 45}, 3, 1479 (1992);
https://doi.org/10.1103/PhysRevA.45.1479.

\bibitem{iran}H.~Fatehizadeh, R.~Gheisari, and H.~Falinejad,
Full calculation of $\mu pd$ and $\mu dt$ muonic bound levels: Combination of Nikiforov-Uvarov
method and variational approach, Ann. Phys. {\bf 385}, 512 (2017);
https://doi.org/10.1016/j.aop.2017.07.017.

\bibitem{varga}K.~Varga and Y.~Suzuki, 
Solution of few-body problems with the stochastic variational method I. Central forces with zero orbital momentum,
Comp. Phys. Comm. {\bf 106}, 157 (1997);
https://doi.org/10.1016/S0010-4655(97)00059-3.

\bibitem{korobov}V.~I.~Korobov, 
Variational Methods in the Quantum Three-Body Problem with Coulomb Interaction,
Phys. Part. Nucl. {\bf 53}, No. 1, 5 (2022);
https://doi.org/10.1134/S1063779622010038.

\bibitem{khan}Md.~A.~Khan, Hyperspherical three-body calculation for exotic atoms,
Few-Body Syst. {\bf 52}: 53-63 (2012);
https://doi.org/10.1007/s00601-011-0264-3.

\bibitem{var}D.~A.~Varshalovich, V.~K.~Khersonsky, E.~V.~Orlenko, and A.~N.~Moskalev,
Quantum theory of angular momentum and its applications, Vol.~I, M., Fizmatlit, 2017.

\bibitem{t4}L.~D.~Landau and E.~M.~Lifshitz, Course of Theoretical
Physics, Vol. 4: Quantum Electrodynamics, Fizmatlit, M., 2008; 
Pergamon, NY, 1977, 3rd ed.

\bibitem{ncm}N.~C.~Mukhopadhyay, Nuclear Muon capture, Phys. Rep. {\bf 30}, No.1, pp.1-144 (1977);
https://doi.org/10.1016/0375-9474(80)90172-4.

\bibitem{scheck}F.~Scheck, Muon Physics, Phys. Rep. {\bf 44}, No.4, pp.187-248 (1978);
https://doi.org/10.1016/0370-1573(78)90014-5.

\bibitem{kaon}C.~Curceanu, C.~Guaraldo, M.~Iliescu, et al., The modern era of light kaonic atom experiments,
Rev. Mod. Phys. {\bf 91}, 025006 (2019);
https://doi.org/10.1103/RevModPhys.91.025006.

\bibitem{asacusa}A.~S\'oter, H.~Aghai-Khozani, D.~Barna, A.~Dax, L.~Venturelli and M.~ Hori,
High-resolution laser resonances of antiprotonic helium in superfluid $^4He$,
Nature {\bf 603}, 411 (2022);
https://doi.org/10.1038/s41586-022-04440-7.

\bibitem{cohen}J.~S.~Cohen, Multielectron effects in capture of antiprotons and muons by helium and neon,
Phys. Rev. A {\bf 62}, 022512 (2000);
https://doi.org/10.1103/PhysRevA.62.022512.

\bibitem{landua}R.~Landua and E.~Klempt, Atomic Cascade of Muonic and Pionic Helium Atoms,
Phys. Rev. Lett. {\bf 48}, 1722 (1982);
https://doi.org/10.1103/PhysRevLett.48.1722.

\bibitem{baird}S.~Baird, C.~J.~Batty, F.~M.~Russell et al., 
Measurements on exotic atoms of helium, Nucl. Phys. A {\bf 392}, 297 (1983);
https://doi.org/10.1016/0375-9474(83)90127-6.

\bibitem{wetmore}R.~J.~Wetmore, D.~C.~Buckle, J.~R.~Kane, and R.~T.~Siegel,
Pionic and muonic X rays in liquid helium, Phys. Rev. Lett. {\bf 19}, 1003 (1967);
https://doi.org/10.1103/PhysRevLett.19.1003.

\bibitem{souder}P.~A.~Souder, D.~E.~Casperson, T.~W.~Crane et al., 
Formation of the Muonic Helium Atom, $\alpha \mu^- e^-$, and Observation of Its Larmor Precession,
Phys. Rev. Lett. {\bf 34}, 1417 (1975);
https://doi.org/10.1103/PhysRevA.22.33.

\bibitem{egger}G.~Backenstoss, J.~Egger, T.~von~Egidy et al., 
Pionic and muonic X-ray transitions in liquid helium, Nucl. Phys. A {\bf 232}, 519 (1974);
https://doi.org/10.1016/0375-9474(74)90637-X.

\bibitem{araki}H. Araki, Quantum-Electrodynamical Corrections to Energy Levels of Helium,
Prog. Theor. Phys. {\bf 17}, 619 (1957);
https://doi.org/10.1143/PTP.17.619.

\bibitem{sucher}J. Sucher, Energy Levels of the Two-Electron Atom to Order $\alpha^3 Ry$; Ionization
Energy of Helium, Phys. Rev. {\bf 109}, 1010 (1958);
https://doi.org/10.1103/PhysRev.109.1010.

\bibitem{korobov2022}V. I. Korobov and J.-Ph. Karr, Spin-orbit interaction in the $HD^+$ ion,
Eur. Phys. J. D {\bf 76}, 197 (2022);
https://doi.org/10.1140/epjd/s10053-022-00522-3.

\bibitem{apm2025}A.~V.~Eskin, V.~I.~Korobov, A.~P.~Martynenko, and F.~A.~Martynenko,
The Energy Spectrum of the Excited States in Mesonic Helium,
Phys. Part. Nucl. {\bf 56}, 2, 235 (2025);
https://doi.org/10.1134/S1063779624701466.


\end{thebibliography}
\end{document}